\documentclass{ieeeaccess}
\usepackage{cite}
\usepackage{amsmath,amssymb,amsfonts}
\usepackage{newtxtext,newtxmath}
\usepackage{enumitem}

\AtBeginDocument{%
	\DeclareSymbolFont{letters}{OML}{ntxmi}{m}{it}%
	\SetSymbolFont{letters}{normal}{OML}{ntxmi}{m}{it}%
	\SetSymbolFont{letters}{bold}{OML}{ntxmi}{b}{it}%
	\DeclareSymbolFont{NewLetters}{OML}{ntxmi}{m}{it}%
	\SetSymbolFont{NewLetters}{normal}{OML}{ntxmi}{m}{it}%
	\SetSymbolFont{NewLetters}{bold}{OML}{ntxmi}{b}{it}%
}

\usepackage{algorithmic}
\usepackage{algorithm}
\usepackage{array}
\usepackage{graphicx}
\usepackage{textcomp}
\usepackage{stfloats}
\usepackage{url}
\usepackage{verbatim}
\usepackage{cite}
\usepackage{diagbox}
\usepackage{color}
\usepackage{soul}
\sethlcolor{white}
\newcommand{\fin}[1]{{\sethlcolor{white}\hl{#1}}}
\def\BibTeX{{\rm B\kern-.05em{\sc i\kern-.025em b}\kern-.08em
		T\kern-.1667em\lower.7ex\hbox{E}\kern-.125emX}}

\begin{document}
	\history{Received 11 June 2026, accepted 20 July 2026.}
	\doi{10.1109/ACCESS.2026.3717843}
	
	\title{Immersive Micromanipulation Integrating Pipette and Injector Operations with McKibben-Based Haptic Sensations for Workload Reduction}
	
	\author{\uppercase{Kenta Yokoe}\authorrefmark{1},~\IEEEmembership{Member,~IEEE},
		\uppercase{Sumiwa Saito}\authorrefmark{2},
		\uppercase{Yuki Funabora}\authorrefmark{3},~\IEEEmembership{Member,~IEEE},
		\uppercase{Tomoko Isomura}\authorrefmark{4},
		and \uppercase{Tadayoshi Aoyama}\authorrefmark{1},~\IEEEmembership{Member,~IEEE}}
	
	\address[1]{Department of Mechanical Systems Engineering, Nagoya University, Nagoya, Aichi, 464-8603, Japan (e-mail: yokoe@nagoya-u.jp, tadayoshi.aoyama@mae.nagoya-u.ac.jp)}
	\address[2]{Department of Micro--Nano Mechanical Science and Engineering, Nagoya University, Nagoya, Aichi, 464-8603, Japan}
	\address[3]{Department of Information and Communication Engineering, Nagoya University, Nagoya, Aichi, 464-8603, Japan}
	\address[4]{Department of Cognitive and Psychological Sciences, Nagoya University, Nagoya, Aichi, 464-8603, Japan}
	% \address[5]{Center for One Medicine Innovative Translational Research, Gifu University, Gifu 501-1193, Japan}
	
	\tfootnote{This work was supported by JST [Moonshot R\&D][Grant Number JPMJMS2214-08], Japan.
		The protocol of this study was approved by the Ethics Committee of the Graduate School of Engineering, Nagoya University (Approval Number: 23-11).
		% --- arXiv posting notice (remove this sentence group if not required) ---
		This is the accepted version of an article published in IEEE Access, vol.~14, pp.~116393--116404, 2026, DOI: 10.1109/ACCESS.2026.3717843. \copyright~2026 The Authors. Open Access under CC BY 4.0 (https://creativecommons.org/licenses/by/4.0/).
		% --- end of arXiv posting notice ---
	}
	
	\markboth
	{Yokoe \headeretal: Immersive Micromanipulation Integrating Pipette and Injector Operations with McKibben-Based Haptic Sensations}
	{Yokoe \headeretal: Immersive Micromanipulation Integrating Pipette and Injector Operations with McKibben-Based Haptic Sensations}
	
	\corresp{Corresponding authors: Kenta Yokoe and Tadayoshi Aoyama (e-mail: yokoe@nagoya-u.jp and tadayoshi.aoyama@mae.nagoya-u.ac.jp).}
	
	\begin{abstract}
		Intracytoplasmic sperm injection (ICSI) requires advanced micromanipulation techniques but relies solely on visual feedback and involves frequent interface switching between pipette movement and injector operations.
		Existing haptic feedback systems primarily focus on pipette puncture forces and do not provide feedback on injector states. 
		We developed an immersive micromanipulation system that unifies operational interfaces and provides McKibben-based haptic sensations to represent aspiration, discharge, and contact between the oocyte and pipette. 
		Users operated both the pipette and injector with a single hand while receiving haptic sensations. 
		A human-participant experiment revealed that the immersive operation interface improved micromanipulation speed and reduced cognitive workload of the micromanipulation compared with conventional methods. 
		Additionally, McKibben-based haptic sensations improved overall system usability. 
		The immersive micromanipulation system with McKibben-based haptic sensations successfully unified operational interfaces and reduced operator workload. 
		%This study presents a practical approach to lowering technical barriers in micromanipulation tasks.
	\end{abstract}	
	
	\begin{keywords}
		Intracytoplasmic sperm injection, haptic sensation, virtual reality, micromanipulation, McKibben artificial muscles
	\end{keywords}
	
	\maketitle
	
	\section{Introduction}
	\label{sec:introduction}
	\IEEEPARstart{M}{icromanipulation} is an essential technique in fields such as biology, medicine, and pharmaceutical science. This technique is employed in various applications, including transgenics, where genes are injected into and manipulated within cells, and in the development of bioabsorbable stents for therapeutic use~\cite{micro1, micro2}. 
	Micromanipulation is also utilized in intracytoplasmic sperm injection (ICSI), an assisted reproductive technology in which a sperm is directly injected into an oocyte. 
	
	\begin{figure}[!t]
		\centering
		\includegraphics[keepaspectratio=true,width=.9\linewidth]{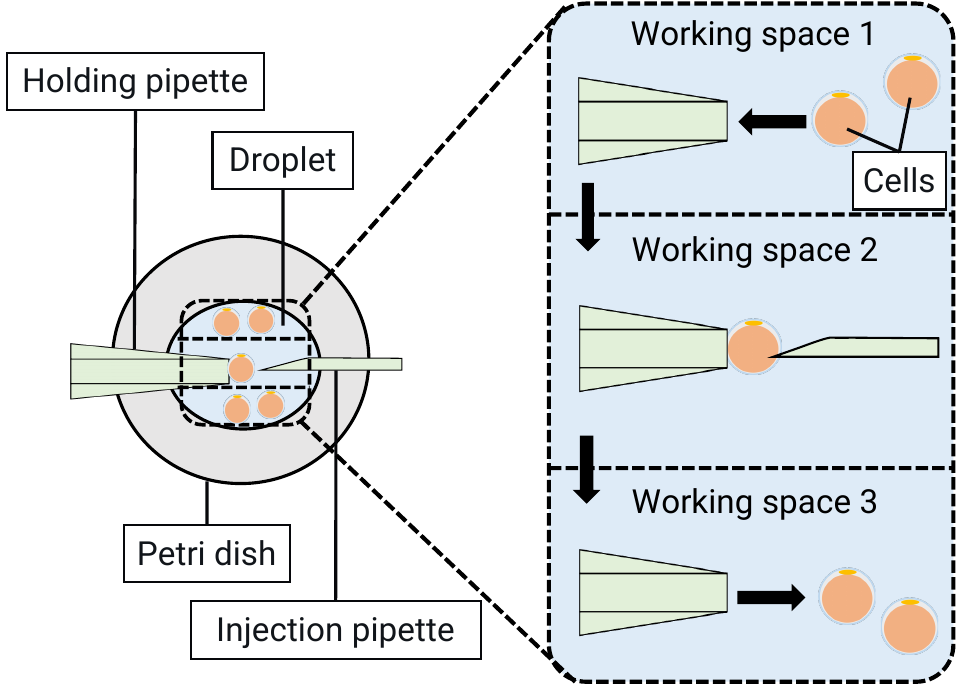}
		\caption{Process of the ICSI procedure showing three working spaces.}
		\label{fig:icsi_process}
	\end{figure}
	
	The basic procedure of ICSI is illustrated in Fig.~\ref{fig:icsi_process}. 
	To prevent the mixing of oocytes, the working space is divided into three sections. 
	Before injection, oocytes are placed in Workspace 1, where they are aspirated and held using a holding pipette and then transferred to Workspace 2. In Workspace 2, the oocytes are rotated so that the polar body is lies at the 12 o'clock or 6 o'clock position, enabling the injection pipette to avoid the spindle during sperm injection. 
	Finally, the oocytes are transferred to Workspace 3 for discharge.
	
	In conventional ICSI environments, operators perform all tasks by relying solely on visual information, which makes depth perception challenging. Moreover, interface switching is required between pipette movement and aspiration/discharge operations. 
	Therefore, ICSI procedures require advanced technical skills and are typically performed by embryologists who specialize in oocyte handling. However, with the recent global increase in demand for ICSI~\cite{icsi_demand}, the shortage of embryologists and their limited experience (with approximately half having less than two years of experience) has become a significant social issue. 
	As a result, systems that support novice ICSI operators are needed.
	
	To address the challenges of depth perception and operational complexity in micromanipulation, we previously developed an immersive micromanipulation system using virtual reality technology~\cite{yokoe2022immersive}. 
	This system integrated a three-dimensional (3D) operational interface for pipette movement. Operators controlled the 3D position of the pipettes through hand tracking in an immersive environment displayed via a head-mounted display (HMD). 
	The system involved an electrically tunable lens for the construction of real-time 3D microscopic images, allowing operators to freely change their viewpoint within the immersive environment. 
	Experimental validation with novice operators demonstrated that the immersive approach significantly improved both the speed and accuracy of pipette positioning compared with conventional two-dimensional microscopy using joystick control~\cite{yokoe2022immersive}.
	However, the previous system had a critical limitation for practical ICSI applications. Aspiration and discharge operations were controlled through separate injectors. Because operators viewed the workspace through an HMD, operators were required to remove the HMD to operate the injectors.
	
	%Conventional ICSI procedures, including our previous system, rely entirely on visual feedback to judge aspiration, discharge, and contact states. However, these operations involve microscale fluid movements and sub-threshold forces that produce only subtle visual cues, making state recognition extremely difficult for novice operators. We hypothesized that integrating aspiration and discharge operations into the unified immersive interface and providing artificial tactile sensations for these difficult-to-perceive states would enable operators to readily acquire operational information, thereby further reducing the technical barriers to performing ICSI.
	
	In this study, we extended our previous work~\cite{yokoe2022immersive} and developed an immersive micromanipulation system by (1) integrating aspiration and discharge operations with pipette movement control into a single hand-tracking interface, (2) developing a wearable fabric actuator based on McKibben artificial muscles to provide haptic sensations representing aspiration and discharge, and (3) delivering contact sensations through another fabric actuator to indicate contact with the oocyte. 
	The proposed system follows the concept of augmented haptics~\cite{saito2024Assistive}, a technology that converts states imperceptible to humans into artificial tactile sensations in human--machine systems. In the proposed system, augmented haptics conveys the aspiration and discharge states of the pipette and the contact between the oocyte and the pipette, which are difficult for operators to perceive during micromanipulation.
	
	\hl{Presenting these states through haptic feedback is expected to reduce an operator's cognitive load. This expectation is based on the Multiple Resource Theory, which states that vision and haptics use separate cognitive resources}~\cite{wickens2002Multiple,wickens2008Multiple}\hl{. In cell injection, which is closely related to ICSI, the micro-scale forces are difficult to perceive, which significantly lowers the success rate of manual operation. Providing force feedback in addition to vision increased the injection success rate compared with vision alone}~\cite{pillarisetti2007Evaluating}\hl{, and a haptic-enabled VR trainer improved the accuracy and success rate of micro-robotic cell injection}~\cite{faroque2018Evaluation}\hl{. The same augmented-haptics approach was applied to microinjection and helped operators, especially beginners, grasp the otherwise imperceptible deformation of the oocyte}~\cite{saito2024Assistive}\hl{. Based on these findings, this study hypothesized that presenting haptic feedback for the aspiration, discharge, and contact states reduces the cognitive load of ICSI-like tasks.}
	
	The key contributions of this work are threefold. 
	First, we demonstrate complete operational unification by controlling pipette movement, aspiration, and discharge through a single hand-tracking interface, thereby eliminating the need for interface switching.
	Second, we develop McKibben-based fabric actuators that provide haptic sensations representing imperceptible states---aspiration, discharge, and contact---which otherwise produce only subtle visual cues.
	Third, we experimentally verify the efficiency of the proposed system in terms of the task completion time and cognitive workload.
	
	The remainder of this paper is organized as follows. 
	Section~\ref{sec:related_work} presents a review of related work on micromanipulation systems. 
	Section~\ref{sec:immersive_micromanipulation_system} describes the proposed system architecture, including the immersive interface and McKibben-based haptic sensations. 
	Section~\ref{sec:experimental_evaluation} presents experimental evaluation in which novice operators performed microbeads manipulation tasks. Section~\ref{sec:discussion} presents a discussion on the implications of the results. 
	Finally, Section~\ref{sec:conclusion} concludes the paper and outlines future work.
	
	\section{Related Work}
	\label{sec:related_work}
	Research on ICSI automation has successfully addressed specific sub-tasks, including cell transfer and puncture~\cite{lu2011Robotica,zhang2019Roboticb}. 
	However, achieving full automation remains challenging due to individual differences in sample properties, such as oocyte size and membrane elasticity, necessitating the continued involvement of human operators for adaptive control~\cite{icsi_automation_problem1}. 
	An alternative approach to address these difficulties is focusing on reducing operator workload by improving the interface instead of pursuing full automation.
	However, operators face high cognitive loads and operational complexity, as operators are required to simultaneously coordinate multiple instruments while relying on limited 2D visual feedback in conventional ICSI setups~\cite{rubino2016ICSIa}. 
	Therefore, systems that integrate these fragmented operations into a unified interface are essential for reducing these technical barriers and supporting human operators.
	
	To achieve such intuitive operation, virtual reality (VR) technology has emerged as a promising solution, enabling the transition from 2D microscopy to immersive 3D visualization. 
	Early VR systems addressed depth perception challenges by providing stereoscopic visualization of the workspace during teleoperation~\cite{ammi2004Virtualized,mehrtash2012Humanassisted,bolopion2013Review}. 
	Recent systems with hand-tracking interfaces allow novice operators to achieve high-speed and accurate micromanipulation comparable to that achieved by experts~\cite{yokoe2022immersive}. 
	However, a critical limitation remains in these immersive systems: aspiration and discharge operations are not integrated into the VR interface, forcing operators to switch devices.
	Furthermore, most of these systems rely entirely on visual feedback and lack haptic information for assessing ICSI states, which produce only subtle visual cues that are difficult to perceive.
	
	To address the limitation of relying solely on visual feedback, haptic feedback methods have been developed to enhance micromanipulation performance.
	Several systems providing haptic feedback during cell membrane puncture have demonstrated improved injection success rates and reduced cell damage~\cite{ammi2005Realistic,pillarisetti2007Evaluating,ghanbari2014Haptic}. 
	However, grounded haptic devices restrict operator movement, making wearable haptic presentation preferable for immersive systems~\cite{pacchierotti2018Steering}.
	Recently, wearable haptic devices using McKibben artificial muscles have been studied~\cite{cacucciolo2020ElectricallyDriven,yu2022selfsensing,yokoe2025Intuitive}.
	These artificial muscles are flexible and provide high haptic output per unit volume compared with other haptic presentation methods~\cite{tondu2012Modelling,yokoe2024Intuitive}; thus, these are suitable for immersive interfaces where users need to move freely.
	
	However, few existing systems integrate an immersive operational interface with McKibben-based wearable haptic feedback to convey imperceptible micromanipulation states in immersive environments.
	In this study, we address these limitations by developing an immersive micromanipulation system that unifies pipette movement, aspiration, and discharge operations in a single hand-tracking interface while providing McKibben-based haptic sensations for these imperceptible states.
	This integration enables operators to perceive critical operational information that was previously inaccessible through visual feedback alone, thereby reducing technical barriers for novice operators.
	
	\section{Immersive Micromanipulation System with McKibben-Based Haptic Sensations}
	\label{sec:immersive_micromanipulation_system}
	\subsection{System Overview}
	\begin{figure}[!t]
		\centering
		\includegraphics[width=\linewidth]{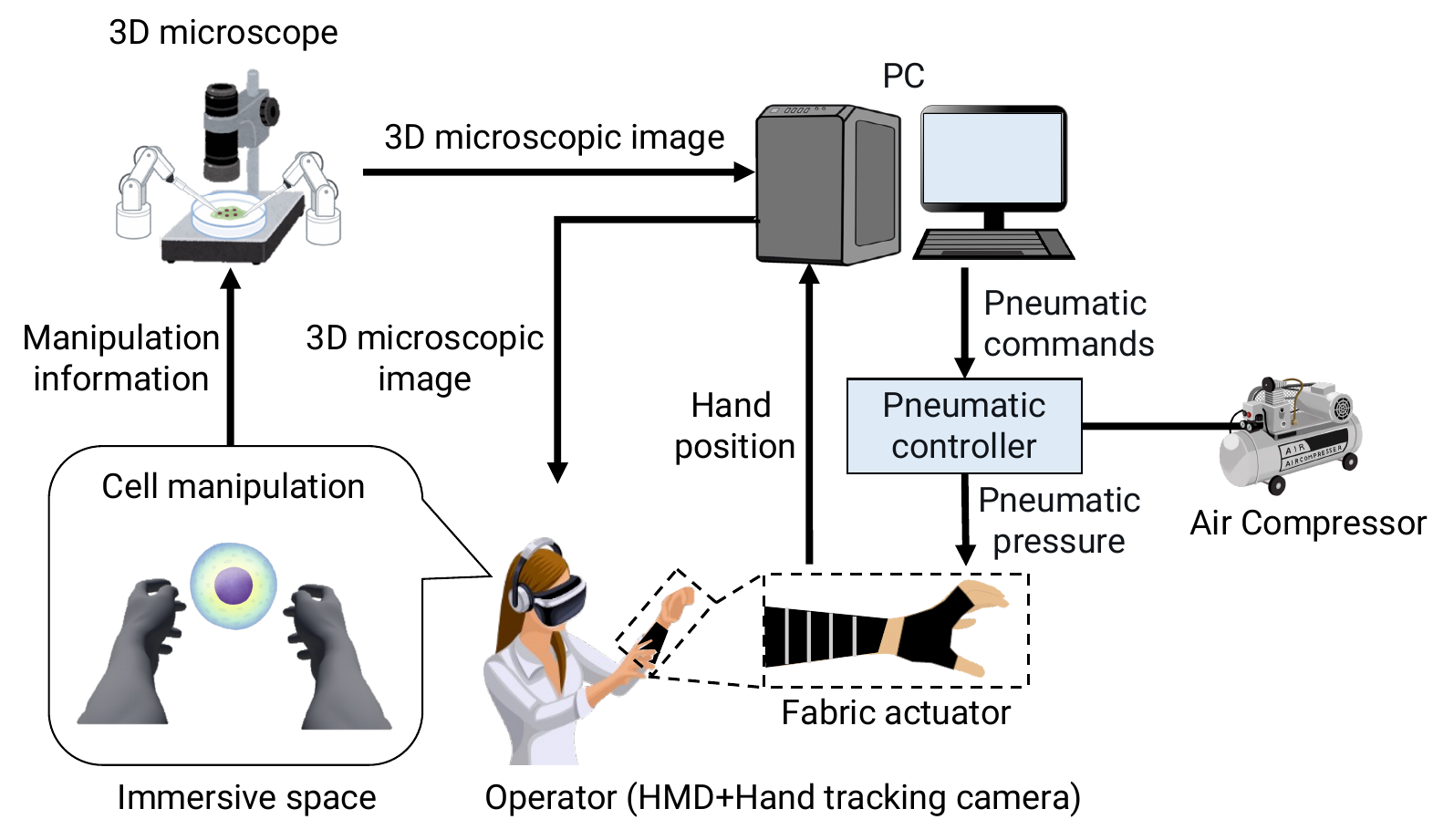}
		\caption{Configuration of the proposed system.}
		\label{fig:system_config}
	\end{figure}
	\begin{figure}[!t]
		\centering
		\includegraphics[width=\linewidth]{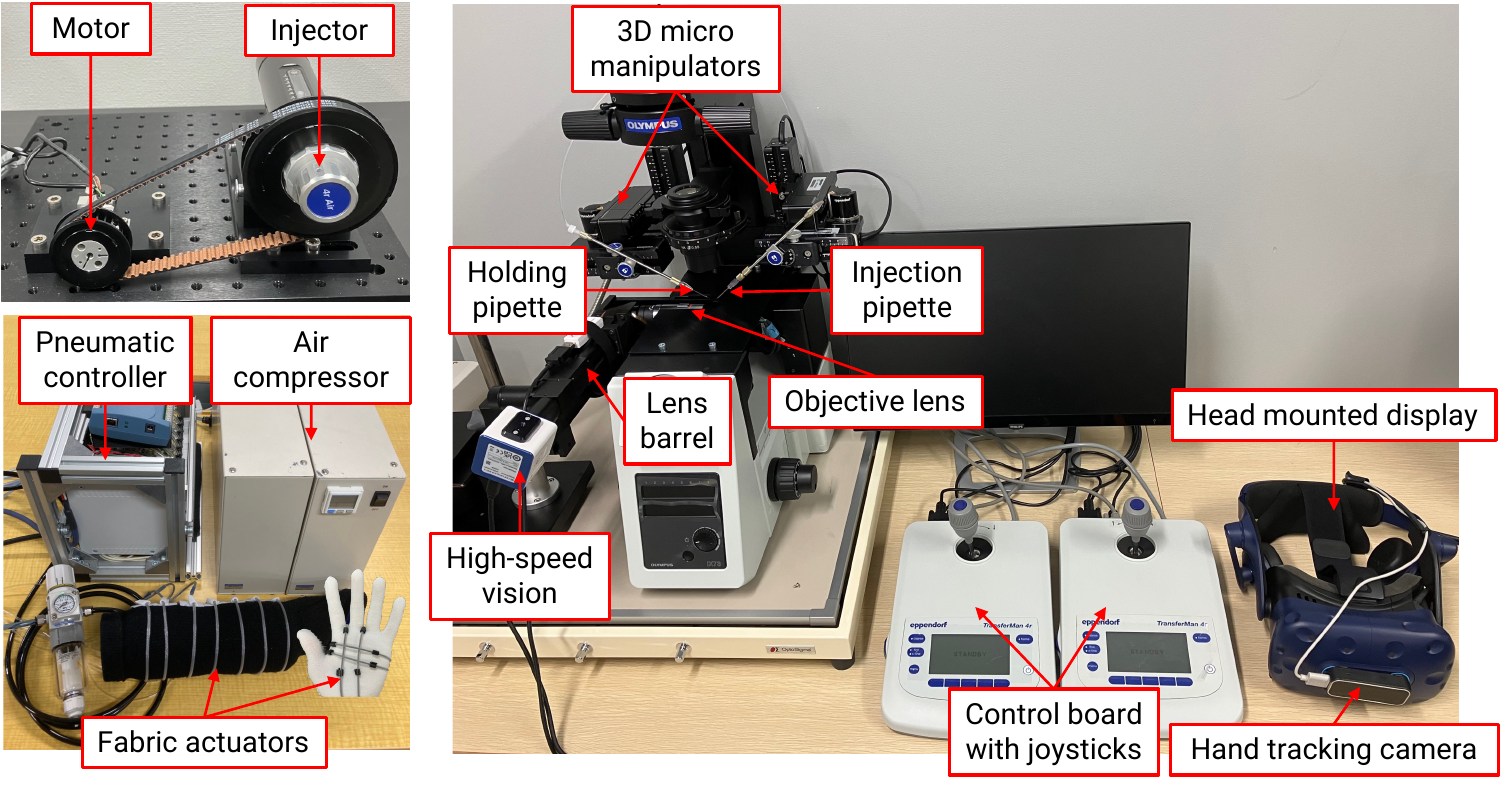}
		\caption{Overview of the proposed system.}
		\label{fig:system_overview}
	\end{figure}
	Figures~\ref{fig:system_config} and \ref{fig:system_overview} illustrate the configuration and overview of the immersive micromanipulation system, featuring McKibben-based haptic sensations. 
	The proposed system comprises two main components: an immersive micromanipulation system and a haptic feedback system.
	The immersive micromanipulation system was based on a previous study~\cite{yokoe2022immersive}.
	The micromanipulation system consists of an inverted microscope (IX73, OLYMPUS), an objective lens (LWD95mm, 10$\times$, Mitutoyo), a high-speed vision system (INFINICAM-UC1, Photron), an electrically tunable lens (ETL) (EL-10-30-C-VIS-LD-MV, Optotune), a lens driver (LensDriver 4, Optotune), a two-axis galvanometer mirror (6210HSM6mm 532 nm, Cambridge Technology), a light source (LA-HDF158AS, Hayashi Repic Corporation), two micromanipulators (TransferMan 4r, Eppendorf), a microinjector (CellTram 4r Air, Eppendorf), a head-mounted display (HMD) (VIVE pro eye, HTC), a hand-tracking camera (LeapMotion, ultraleap), a computer (OS Windows 10 Home 64bit, CPU Intel (R) Core (TM) i9-9900KF 3.60 GHz, memory 32GB RAM, GPU NVIDIA GeForce RTX 2080 SUPER), and a D/A board (PCX-340416 Interface). 
	In addition, the micromanipulation system employs a servomotor (RSF-5B100-E050-C, Harmonic Drive Systems), servo pack (HA-680-4B-24, Harmonic Drive Systems), timing belt (HTBN475S5M-100, MISUMI), and pulleys (HTPB27S5M100-A-P21, HTPB49SM100-B-P41, MISUMI) to control the injector.
	The magnification of the objective lens is five$\times$, with a working distance of 34 mm. The pixel pitch of the high-speed vision system is 10.0 $\mu$m. 
	The focal-length range of the ETL is 100--200 mm at 30$^\circ$C.
	The immersive micromanipulation system creates an immersive micromanipulation environment using the method described in~\cite{yokoe2022immersive}.
	The user of the proposed system is immersed in an immersive micromanipulation environment through an HMD and manipulates virtual objects, which are used to control the holding pipette and injector in real-time. 
	
	The haptic feedback system consists of an air compressor (ACP-39SLA, TAKAGI Corporation), programmable logic controllers (M-DUINO PLC 54ARA I/Os Analog/Digital Plus, Industrial Shields), electropneumatic regulators (CRCB-0135W, Koganei Corporation), and fabric actuators.
	The fabric actuators are composed of McKibben artificial muscles (EM20, S-muscle), a cotton arm cover (Takeshikun Arm Cover 30, Mie Chemical Industry), and a cotton glove (antibacterial and antiviral glove N-3800, CO-COS NOBUOKA).
	These actuators deliver haptic sensations to the operator's left arm and hand.
	The arm receives haptic feedback representing aspiration and discharge of fluid through the holding pipette, while the hand receives haptic feedback representing contact between the oocyte and holding pipette.
	These haptic sensations help users judge fluid flow and contact between the holding pipette and the oocyte, which could previously only be assessed visually.
	
	\subsection{Immersive Micromanipulation System to Operate Holding Pipette and Injector}
	In the proposed system, the position of the holding pipette and the rotational operation of the injector are controlled using hand gestures.
	The holding pipette and injector were operated as follows:
	\begin{enumerate}%[label=(\alph*)]
		\item The operator moves the left hand within the immersive micromanipulation environment.
		\item A hand-tracking camera captures the displacement and fingertip angle of the left hand.
		\item The actual holding pipette is moved according to the obtained displacement of the left hand.
		\item The aspiration and discharge operations of the actual injector are controlled based on the fingertip angle.
		%\item A 3D image is presented to the operator in the immersive micromanipulation environment.
	\end{enumerate}
	For the pipette movement operation, the displacement, $[x_{\mathrm{m}}, y_{\mathrm{m}}, z_{\mathrm{m}}]^\top$, of the left hand is obtained and used to move the pipette. 
	The system calculates the displacement between the initial hand position, $(x_{\mathrm{h}}^0, y_{\mathrm{h}}^0, z_{\mathrm{h}}^0)^\top$, and the current hand position, $(x_{\mathrm{h}}, y_{\mathrm{h}}, z_{\mathrm{h}})^\top$, and the pipette is controlled according to this displacement as follows:
	\begin{equation}
		\begin{pmatrix}
			x_{\mathrm{p}} \\
			y_{\mathrm{p}} \\
			z_{\mathrm{p}}
		\end{pmatrix} 
		=
		\begin{pmatrix}
			x_{\mathrm{h}}-x_{\mathrm{h}}^0 \\
			y_{\mathrm{h}}-y_{\mathrm{h}}^0 \\
			z_{\mathrm{h}}-z_{\mathrm{h}}^0
		\end{pmatrix},
		\label{eq:movement}
	\end{equation}
	where $x_{\mathrm{p}}, y_{\mathrm{p}}, z_{\mathrm{p}}$ denotes the position of the holding pipette. 
	
	\begin{figure}[!t]
		\centering
		\includegraphics[width=\linewidth]{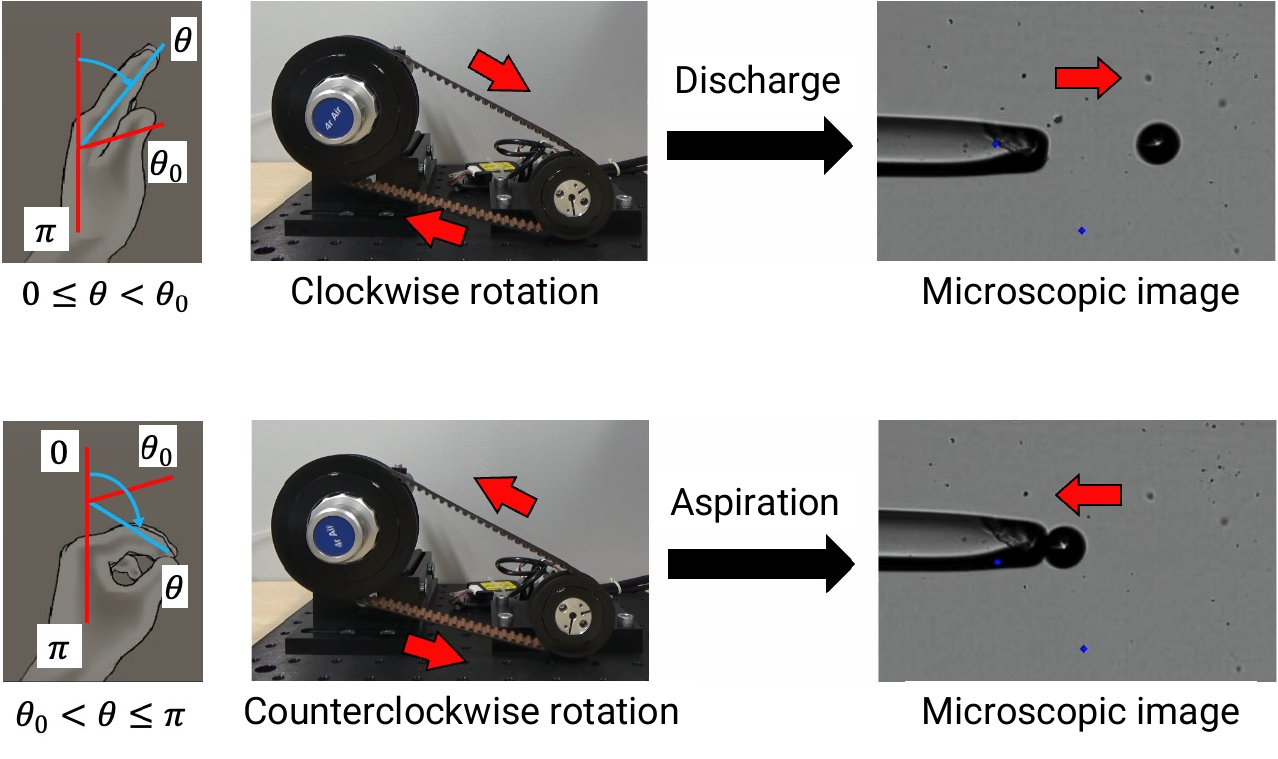}
		\caption{Motor movement according to fingertip angle.}
		\label{fig:rotation}
	\end{figure}
	The system drives the motor to rotate the injector based on the fingertip angle, as shown in Fig.~\ref{fig:rotation}. 
	The target motor angle, $R$, is determined as follows:
	\begin{equation}
		R=G(\theta-\theta_0),
		\label{eq:target_angle}
	\end{equation}
	where $\theta$ is the fingertip angle, $\theta_0$ is the reference angle, and $G$ is a constant.
	As shown in Fig.~\ref{fig:rotation}, when the fingertip angle, $\theta$, is smaller than the reference angle, $\theta_0$, the motor rotates clockwise to perform the discharge operation. When $\theta$ is larger than the reference angle, $\theta_0$, the motor rotates counterclockwise to perform the aspiration operation.
	
	\subsection{Haptic Sensations to Present Aspiration/Discharge and Contact}
	\begin{figure}[!t]
		\centering
		\includegraphics[width=.8\linewidth]{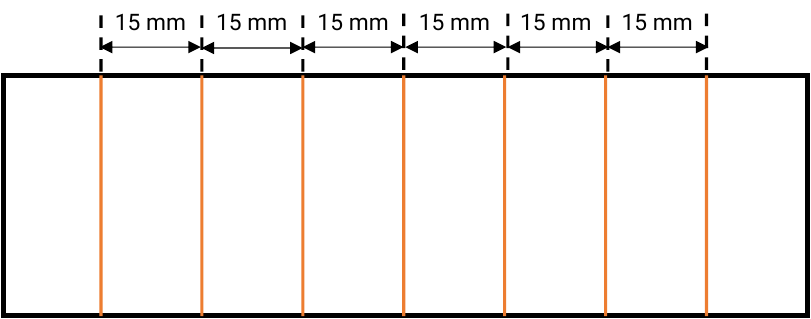}
		\caption{Artificial muscle arrangement of the fabric actuator for providing aspiration and discharge sensations. Orange lines represent the artificial muscles. Square box represents the cotton arm cover.}
		\label{fig:arrange_arm}
	\end{figure}
	\begin{figure}[!t]
		\centering
		\includegraphics[width=.85\linewidth]{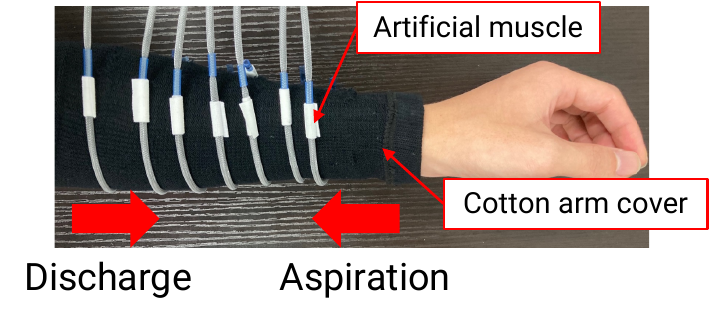}
		\caption{Fabric actuator for providing aspiration and discharge sensations.}
		\label{fig:arm}
	\end{figure}
	Figures~\ref{fig:arrange_arm} and \ref{fig:arm} show the artificial muscle arrangement and an overview of the arm-type fabric actuator that provides haptic sensations to the arm to perform aspiration and discharge operations, respectively. 
	The actuator consists of seven artificial muscles attached to a cotton arm cover. 
	The seven artificial muscles are wrapped around the arm and spaced at 15 mm intervals. This spacing is designed to be below the two-point discrimination threshold of the forearm in individuals aged 18 to 28 years \cite{stevens1996Spatial}. Therefore, users find it difficult to distinguish which of the two adjacent artificial muscles contract. This design enables users to perceive the sensation of fluid flowing along the arm as the artificial muscles contract sequentially from either the wrist or the elbow.
	The applied pneumatic pressure of all seven artificial muscles, $P_{\mathrm{a}}$, is proportional to the rotational speed, $\dot{R}$, of the motor, expressed as follows:
	\begin{equation}
		P_{\mathrm{a}}=a\dot{R},
		\label{eq:air_aspiration}
	\end{equation}
	where $a$ is a constant. \fin{The pressure $P_{\mathrm{a}}$ is in kPa and the rotational speed $\dot{R}$ is in rev/s; therefore, $a$ has the unit of kPa$\cdot$s/rev.}
	During aspiration, pneumatic pressure is applied sequentially at 100 ms intervals, starting from the artificial muscles near the wrist. During discharge, pressure is applied at 100 ms intervals starting from the muscles near the elbow, thereby providing the sensation of aspiration and discharge.
	
	\begin{figure}[!t]
		\centering
		\includegraphics[width=\linewidth]{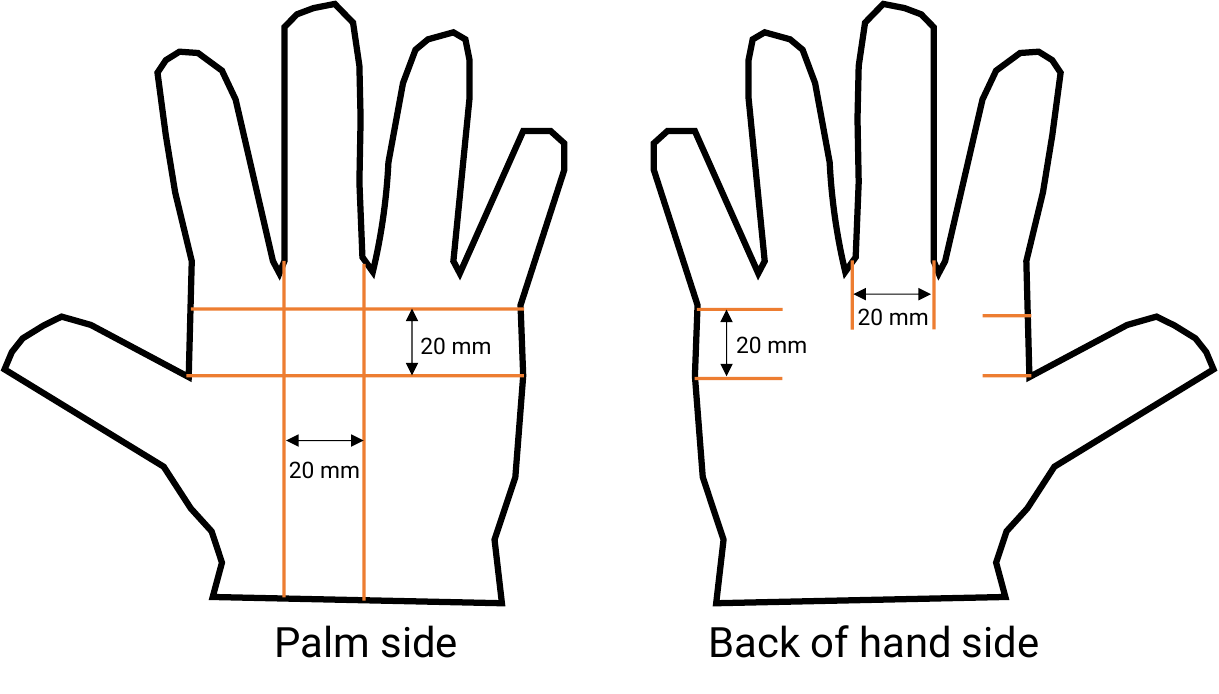}
		\caption{Artificial muscle arrangement of the fabric actuator for providing contact sensations. Orange lines represent the artificial muscles.}
		\label{fig:arrange_glove}
	\end{figure}
	\begin{figure}[!t]
		\centering
		\includegraphics[width=\linewidth]{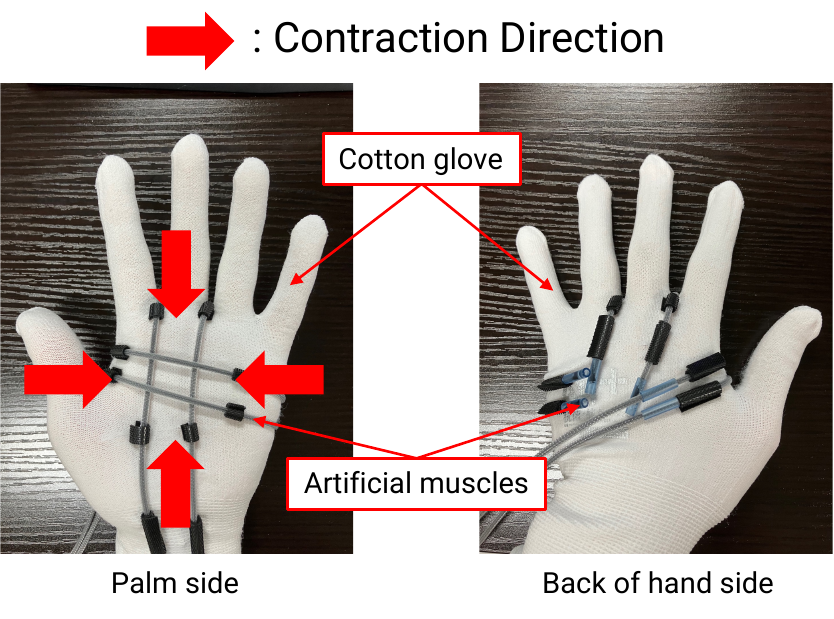}
		\caption{Fabric actuator for providing contact sensations.}
		\label{fig:glove}
	\end{figure}
	Figures~\ref{fig:arrange_glove} and \ref{fig:glove} show the artificial muscle arrangement and overview of the glove-type fabric actuator that provides haptic sensations to the hand for detecting contact between the holding pipette and the oocyte. 
	The actuator consists of four artificial muscles attached to a cotton glove. 
	One artificial muscle extends from the wrist on the palm, passes between the index and middle fingers, and reaches the back of the hand. A parallel muscle extends from the wrist to the palm, passing between the middle and ring fingers to the back of the hand.
	The remaining two artificial muscles run parallel to each other, passing between the thumb and index fingers and beneath the little finger.
	Contact sensation is provided by applying a constant pneumatic pressure, $P_{\mathrm{h}}$, whenever contact between the holding pipette and the oocyte is detected. 
	The proposed system presents contact sensation to the user by applying a predetermined pneumatic pressure, $P_{\mathrm{h}}$, to the four artificial muscles when the distance between the oocyte center and the holding pipette tip falls below a threshold distance.
	\hl{The red arrows in Fig.}~\ref{fig:glove}\hl{ indicate the contraction direction of each artificial muscle. The four artificial muscles contract along these directions, and the resulting contraction compresses the hand to create a sensation of gripping across the palm. The artificial muscles are arranged in this manner to produce the gripping sensation across the entire hand, and this sensation represents the contact between the holding pipette and the oocyte.}
	
	\hl{The thin, soft fabric actuators are lightweight, generate no heat, and can be worn on the hand and arm for long periods. The continuous contraction force of the muscles represents the sustained contact between the holding pipette and the oocyte. This continuous pressure stimulates slowly adapting mechanoreceptors, which keep firing while the skin stays indented, and therefore, suit the sustained contact sensation}~\cite{johnson2001Roles}\hl{. Vibration and electrical stimulation activate rapidly adapting receptors, which are desensitized within a short time and weaken the perceived intensity}~\cite{bensmaia2005Vibratory}\hl{. Without an air compressor, multiple vibration motors could also present the aspiration and discharge sensations as apparent motion. However, the air compressor is placed away from the operator and connected through thin air tubes to prevent it from restricting the operator's motion. The more severe constraint for the immersive interface is interference with the vision-based hand tracking. The thin and soft fabric actuators, unlike vibration motors mounted on the hand, do not obstruct hand tracking.}
	
	\hl{The proposed system provides haptic feedback based on the distance between the holding pipette and the oocyte; therefore, the feedback speed and accuracy affect the operation. The high-speed vision detects contact within a few milliseconds.}\fin{ The accuracy of the contact detection is limited by the imaging resolution of the high-speed vision system.}\hl{ The pneumatic McKibben artificial muscles have a delay of approximately $0.1$ s from the command to the pressure change and a rise time of approximately $1.0$ s}~\cite{peng2023FunabotSuit}\hl{. As reported in the previous study, kinesthetic perception requires a sufficient stimulus duration, rather than a high response speed. The aspiration, discharge, and contact states in micromanipulation are sustained rather than instantaneous. Therefore, this response time does not impair the perception of these states. Regarding the magnitude of the feedback, the proposed system follows the concept of augmented haptics}~\cite{saito2024Assistive}\hl{ and amplifies the imperceptible small sensations into perceivable tactile sensations. Thus, the feedback is designed to emphasize these states, rather than to underrepresent the actual force. Furthermore, because the McKibben actuators are based on soft robotics, they do not forcibly move the operator's hand. Rather, they simply present subtle forces to the operator. The operator, therefore, always performs the operation actively and retains full control, which prevents accidental excessive pressure and allows the operator to stop the operation immediately if necessary. In addition, the mechanical structure of the injector and a software threshold on the motor limit the actual injection pressure. An error in the haptic feedback, therefore, does not directly apply excessive pressure.}
	
	\section{Evaluation of the Proposed System}
	\label{sec:experimental_evaluation}
	\subsection{Experimental Setup}
	\begin{figure}[!t]
		\centering
		\includegraphics[keepaspectratio=true,width=0.6\linewidth]{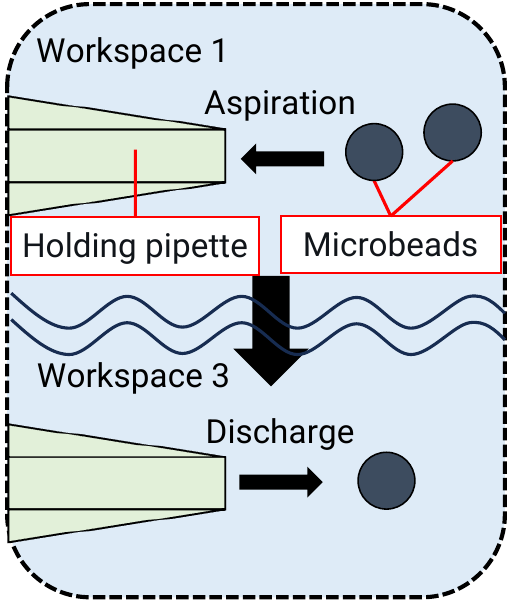}
		\caption{Experimental task based on the actual ICSI procedure.}
		\label{fig:experiment3_1}
	\end{figure}
	We conducted an experiment with human participants to evaluate the effectiveness of the proposed system. 
	Six participants with no prior experience in micromanipulation were recruited for the experiment.
	\hl{The number of participants is comparable to that in previous studies on immersive micromanipulation}~\cite{yokoe2022immersive}\hl{.}
	\fin{The within-subject design, in which every participant performed all five conditions, also controls individual differences and attains sufficient statistical power with fewer participants than a between-subject design.}
	Informed consent was obtained from all participants prior to their participation in the experiment.
	\hl{This study was approved by the Ethics Committee of the Graduate School of Engineering, Nagoya University (Approval Number: 23-11).}
	Figure~\ref{fig:experiment3_1} illustrates the experimental task procedure, which was modeled after an actual ICSI procedure. 
	As described in Section~\ref{sec:introduction}, the oocyte is aspirated in Workspace 1 and discharged in Workspace 3 during the actual ICSI procedure. 
	The experimental task involved aspirating two microbeads from Workspace 1, transferring them to Workspace 3, and discharging them there.
	\hl{The task used microbeads instead of oocytes, which vary between individuals. The microbeads made it possible to measure the inherent capability of the system without the influence of sample variation.}
	Each participant completed the experimental task under five conditions.
	\begin{enumerate}[label=(\Alph*)]
		\item Conventional micromanipulation system,
		\item Immersive micromanipulation system without any haptic sensation,
		\item Immersive micromanipulation system with aspiration/discharge sensations,
		\item Immersive micromanipulation system with contact sensations,
		\item Immersive micromanipulation system with both aspiration/discharge and contact sensations.
	\end{enumerate}
	Each participant performed the experimental task three times per condition. 
	Before the experimental task, participants were instructed on how to perform the task and operate the micromanipulation systems. 
	\hl{Each participant practiced the joystick operation for 10 min before performing Condition (A) for the first time and practiced the immersive operation for 10 min before performing the immersive Conditions (B)--(E) for the first time. The participants needed no extra time to become accustomed to each haptic condition because the haptic feedback used in this study was simple.}
	After the practice, participants performed the experimental task.
	The order of experimental conditions was counterbalanced for each participant.
	\hl{In the experiment, the constant $G$ in Equation~(}\ref{eq:target_angle}\hl{) was set to approximately $0.060$ rev/deg, derived from the motor-to-injector gear ratio of $1.8148$ and a design in which a $30^\circ$ grab angle corresponds to one injector revolution. The constant $a$ in Equation~(}\ref{eq:air_aspiration}\hl{) was set to $200$ }\fin{kPa$\cdot$s/rev}\hl{ such that the applied pneumatic pressure spanned the range of the actuator over the typical motor speeds, with the pressure limited to $400$ kPa.}
	
	Evaluation metrics included task completion time, NASA-TLX WWL Score, and subjective evaluation through questionnaires.
	The questionnaire was administered to evaluate the usability of the proposed system using a seven-point Likert scale, \hl{where a score of 1 indicated the most difficult and a score of 7 indicated the easiest,} for the following four questions:
	\begin{enumerate}[label=Q\arabic*:]
		\item Ease of pipette movement
		\item Ease of aspiration/discharge
		\item Ease of holding a bead
		\item Ease of completing the task
	\end{enumerate}
	\hl{In clinical micromanipulation such as ICSI, the operation time is an important indicator of operation quality. A long operation keeps the oocyte outside the incubator and exposes it to temperature changes and atmospheric oxygen. The oxygen exposure induces oxidative stress and reduces developmental competence}~\cite{wale2016Effects}\hl{. Therefore, clinical guidelines recommend keeping the injection procedure short to prevent oocyte damage}~\cite{eshre2016Revised,simopoulou2016Making}\hl{. In the experimental task, a failed aspiration or discharge or a dropped microbead required the operator to repeat the operation, which increased the completion time; therefore, the operation time may also partly reflect the possibility of operational errors.}
	
	\hl{All evaluation metrics were analyzed using the Friedman test} \cite{friedman}\hl{, and its effect size was reported as Kendall's coefficient of concordance $W$, which ranges from 0 to 1. With six participants, the assumptions of normality and sphericity cannot be verified reliably. Therefore, the Friedman test was adopted as a non-parametric test for within-subject comparison across all metrics. The task completion time was recorded for all three trials of each participant, which gives 18 measurements per condition. The NASA-TLX WWL and the questionnaire items yielded a single value per participant, which gave six measurements per condition. When the Friedman test indicated a significant difference, post-hoc pairwise comparisons were performed using Dunn's test}~\cite{wilcoxon}\hl{ with Bonferroni correction for the number of pairwise comparisons. This procedure was implemented in the related-samples non-parametric Friedman test of SPSS. The Bonferroni-corrected $p$-value and the difference in mean ranks are reported for each comparison.}
	
	\subsection{Experimental Results}
	\begin{figure}[!t]
		\centering
		\includegraphics[keepaspectratio=true,width=\linewidth]{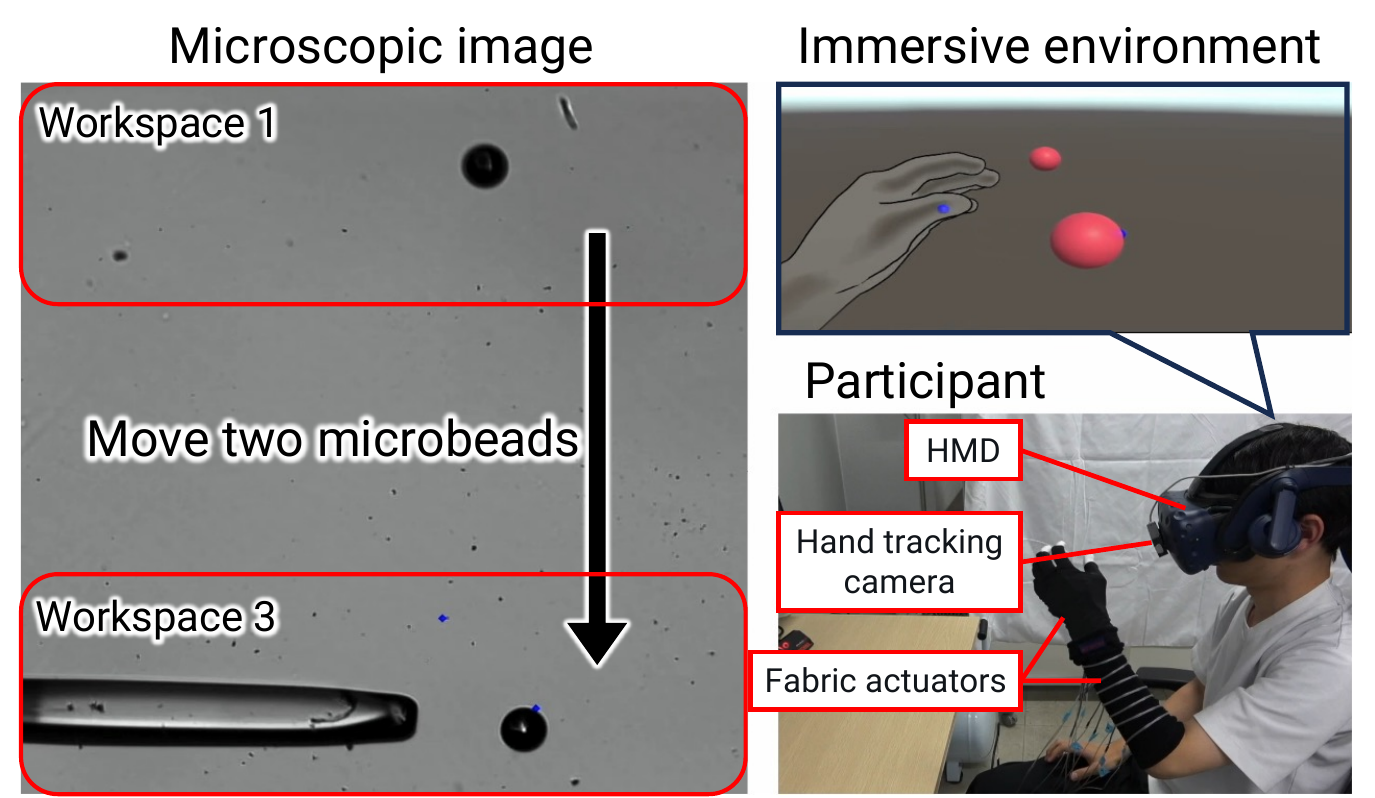}
		\caption{Experimental scene of manipulation using microbeads.}
		\label{fig:experiment3_2}
	\end{figure}
	Figure~\ref{fig:experiment3_2} shows a scene where a participant conducted the experimental task under Condition (E), using the immersive micromanipulation system with both aspiration/discharge and contact sensations.
	Participants manipulated the holding pipette and injector indirectly by moving their hands to transfer the microbead while observing only the immersive environment.
	
	\begin{figure}[!t]
		\centering
		\includegraphics[keepaspectratio=true,width=\linewidth]{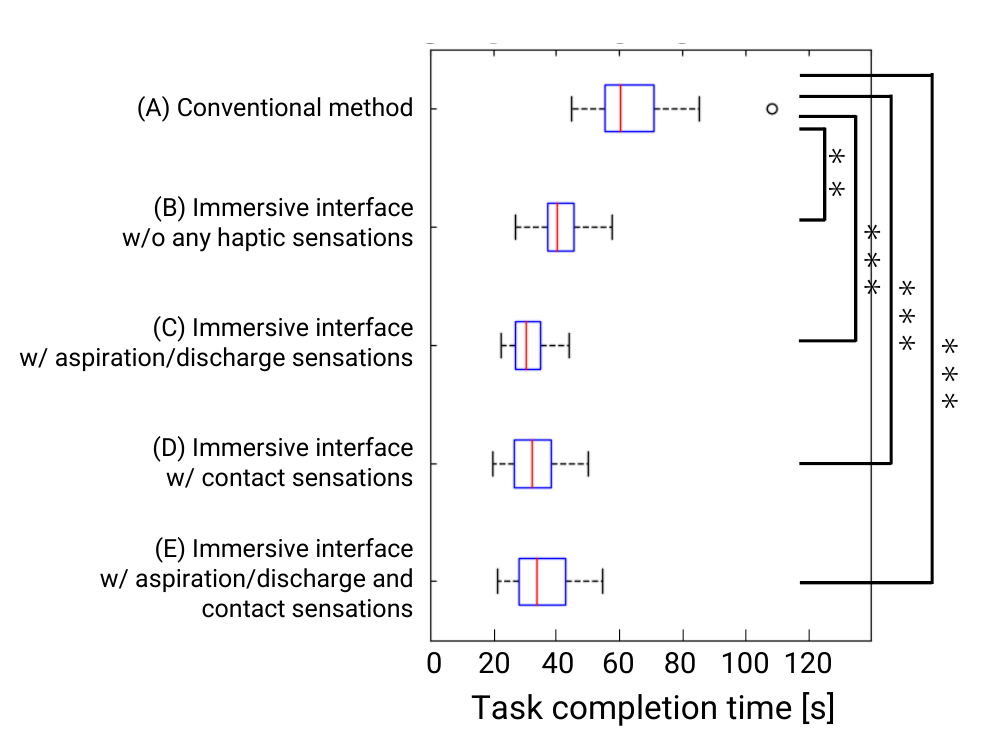}
		\caption{Boxplot of task completion time, showing significant improvement with the immersive VR system. \hl{The asterisks indicate the Bonferroni-corrected significance level of Dunn's pairwise comparison: $^{*}$~$p<0.05$, $^{**}$~$p<0.01$, and $^{***}$~$p<0.001$.}}
		\label{fig:result_time}
	\end{figure}
	\begin{table*}[!t]
		\caption{Experimental result of task completion time [s].}
		\label{table:result_time}
		\centering
		\begin{tabular}{c|ccccc}
			\hline
			Condition & (A) & (B) & (C) & (D) & (E) \\
			Immersive system & --- & \checkmark & \checkmark & \checkmark & \checkmark \\
			Aspiration/discharge sensation & --- & --- & \checkmark & --- & \checkmark \\
			Contact sensation & --- & --- & --- & \checkmark & \checkmark \\ \hline
			Participant (a) & 92.6 & 39.4 & 32.3 & 24.3 & 22.2\\ 
			Participant (b) & 64.3 & 35.6 & 36.3 & 36.9 & 42.9\\ 
			Participant (c) & 60.0 & 45.9 & 34.8 & 37.3& 35.3\\ 
			Participant (d) & 59.7 & 38.5 & 30.5 & 38.2 & 32.1\\ 
			Participant (e) & 53.4 & 35.4 & 24.6 & 26.5 & 40.3\\ 
			Participant (f) & 63.1 & 47.9 & 28.3 & 33.9 & 38.3\\ \hline
			Median & 61.5 & 38.9$^{**}$ & 31.4$^{***}$ & 35.4$^{***}$ & 36.8$^{***}$\\ \hline
		\end{tabular}
		
		\vspace{2mm}
		\footnotesize{\hl{Using Dunn's pairwise comparison with Bonferroni correction for all 18 trials, a significant difference is observed upon comparison of Condition (A) with all other conditions; ($^{**}$~$p<0.01$, $^{***}$~$p<0.001$).}}
	\end{table*}
	Table~\ref{table:result_time} lists the \hl{mean} task completion times for each participant under each condition, and Figure~\ref{fig:result_time} shows \hl{the box plots of all 18 trials per condition (six participants $\times$ three trials)}.
	The median task completion time across participants was approximately \hl{61.5} s under the conventional micromanipulation system (condition (A)) and 30--40 s under the immersive micromanipulation system (conditions (B)--(E)).
	\hl{The statistical analysis used these 18 measurements per condition. The Friedman test revealed a significant difference among the conditions ($p < 0.001$, $W = 0.59$). Post-hoc Dunn's pairwise comparisons with Bonferroni correction showed that the conventional micromanipulation system (Condition (A)) was significantly slower than each of the immersive micromanipulation systems (Conditions (B)--(E)). The Bonferroni-corrected $p$-value was $0.007$ for the comparison with Condition (B) and below $0.001$ for Conditions (C), (D), and (E).}
	\fin{The three trials of each participant were treated as independent blocks, and the nesting of the trials within each participant was not modeled, which is a limitation of the present analysis.}
	These results confirm that the immersive micromanipulation system improves oocyte manipulation speed.
	However, the presence of haptic sensations did not significantly affect micromanipulation speed.
	
	\begin{table}[!t]
		\caption{Experimental result of NASA-TLX WWL Score.}
		\label{table:result_NASA}
		\centering
		\begin{tabular}{c|cccc}
			\hline
			Condition & (B) & (C) & (D) & (E) \\
			Aspiration/discharge sensation & --- & \checkmark & --- & \checkmark \\
			Contact sensation & --- & --- & \checkmark & \checkmark \\ \hline
			Participant (a) & 42.0 & 50.0 & 43.3 & 40.7 \\ 
			Participant (b) & 56.7 & 49.7 & 54.7 & 46.7\\ 
			Participant (c) & 57.3 & 49.3 & 51.3 & 47.3\\ 
			Participant (d) & 61.7 & 38.7 & 44.3 & 44.7\\ 
			Participant (e) & 80.3 & 28.3 & 64.7 & 24.3 \\
			Participant (f) & 82.0 & 50.0 & 50.0 & 48.7\\ \hline
			Median & 59.5 & 49.5 & 50.7 & 45.7$^{*}$\\ \hline
		\end{tabular}
		
		\vspace{2mm}
		\footnotesize{\hl{A significant difference is observed compared with Condition (B) by Dunn's pairwise comparison with Bonferroni correction ($^{*}$~$p<0.05$).}}
	\end{table}
	Table~\ref{table:result_NASA} shows the NASA-TLX WWL for each participant under conditions (B)--(E).
	\hl{The Friedman test among conditions (B)--(E) revealed a significant difference ($p = 0.015$, $W = 0.58$). A post-hoc Dunn's pairwise comparison with Bonferroni correction showed a significant difference between Condition (B) (without any haptic sensations) and condition (E) (with aspiration/discharge and contact sensations) ($p = 0.010$, difference in mean ranks $= 2.33$). By contrast, no significant differences were found among the other conditions.}
	\hl{Table~}\ref{table:result_subscale}\hl{ shows the six NASA-TLX subscale ratings of each participant under Conditions (B)--(E). Among the subscales, the mental demand decreased the most from Condition (B) to Condition (E), with the median decreasing from $65.0$ to $27.5$. The physical demand changed little, with the median changing from $57.5$ to $55.0$. This result indicates that the workload reduction was mainly cognitive rather than physical.}
	
	\begin{table*}[!t]
		\caption{\hl{NASA-TLX subscale ratings of each participant under Conditions (B)--(E) (0--100 scale; a higher score indicates a higher demand).}}
		\label{table:result_subscale}
		\centering
		\begin{tabular}{cc|cccccc}
			\hline
			Participant & Condition & \hl{Mental Demand} & \hl{Physical Demand} & \hl{Temporal Demand} & \hl{Performance} & \hl{Effort} & \hl{Frustration} \\ \hline
			\hl{(a)} & \hl{(B)} & \hl{60} & \hl{40} & \hl{20} & \hl{50} & \hl{30} & \hl{40} \\
			& \hl{(C)} & \hl{20} & \hl{50} & \hl{20} & \hl{60} & \hl{50} & \hl{40} \\
			& \hl{(D)} & \hl{10} & \hl{50} & \hl{20} & \hl{60} & \hl{30} & \hl{20} \\
			& \hl{(E)} & \hl{10} & \hl{50} & \hl{20} & \hl{60} & \hl{30} & \hl{20} \\ \hline
			\hl{(b)} & \hl{(B)} & \hl{70} & \hl{70} & \hl{20} & \hl{30} & \hl{60} & \hl{40} \\
			& \hl{(C)} & \hl{40} & \hl{60} & \hl{20} & \hl{45} & \hl{60} & \hl{40} \\
			& \hl{(D)} & \hl{70} & \hl{60} & \hl{20} & \hl{50} & \hl{60} & \hl{30} \\
			& \hl{(E)} & \hl{20} & \hl{60} & \hl{20} & \hl{80} & \hl{40} & \hl{20} \\ \hline
			\hl{(c)} & \hl{(B)} & \hl{20} & \hl{40} & \hl{50} & \hl{70} & \hl{30} & \hl{30} \\
			& \hl{(C)} & \hl{40} & \hl{60} & \hl{50} & \hl{30} & \hl{70} & \hl{60} \\
			& \hl{(D)} & \hl{70} & \hl{60} & \hl{50} & \hl{30} & \hl{60} & \hl{50} \\
			& \hl{(E)} & \hl{30} & \hl{60} & \hl{50} & \hl{40} & \hl{70} & \hl{60} \\ \hline
			\hl{(d)} & \hl{(B)} & \hl{60} & \hl{45} & \hl{40} & \hl{30} & \hl{60} & \hl{80} \\
			& \hl{(C)} & \hl{40} & \hl{30} & \hl{50} & \hl{45} & \hl{40} & \hl{45} \\
			& \hl{(D)} & \hl{20} & \hl{50} & \hl{50} & \hl{70} & \hl{10} & \hl{55} \\
			& \hl{(E)} & \hl{30} & \hl{45} & \hl{50} & \hl{70} & \hl{55} & \hl{45} \\ \hline
			\hl{(e)} & \hl{(B)} & \hl{80} & \hl{85} & \hl{65} & \hl{75} & \hl{85} & \hl{75} \\
			& \hl{(C)} & \hl{10} & \hl{10} & \hl{30} & \hl{55} & \hl{30} & \hl{10} \\
			& \hl{(D)} & \hl{85} & \hl{70} & \hl{30} & \hl{50} & \hl{70} & \hl{65} \\
			& \hl{(E)} & \hl{25} & \hl{20} & \hl{10} & \hl{35} & \hl{35} & \hl{25} \\ \hline
			\hl{(f)} & \hl{(B)} & \hl{90} & \hl{80} & \hl{20} & \hl{60} & \hl{85} & \hl{70} \\
			& \hl{(C)} & \hl{50} & \hl{60} & \hl{20} & \hl{60} & \hl{30} & \hl{30} \\
			& \hl{(D)} & \hl{60} & \hl{70} & \hl{20} & \hl{60} & \hl{20} & \hl{20} \\
			& \hl{(E)} & \hl{40} & \hl{60} & \hl{30} & \hl{50} & \hl{40} & \hl{60} \\ \hline
			\hl{Median} & \hl{(B)} & \hl{65} & \hl{57.5} & \hl{30} & \hl{55} & \hl{60} & \hl{55} \\
			& \hl{(C)} & \hl{40} & \hl{55} & \hl{25} & \hl{50} & \hl{45} & \hl{40} \\
			& \hl{(D)} & \hl{65} & \hl{60} & \hl{25} & \hl{55} & \hl{45} & \hl{40} \\
			& \hl{(E)} & \hl{27.5} & \hl{55} & \hl{25} & \hl{55} & \hl{40} & \hl{35} \\ \hline
		\end{tabular}
		
		\vspace{2mm}
		\footnotesize{\hl{Condition (B) provides no haptic sensation. Condition (C) provides aspiration and discharge sensations. Condition (D) provides contact sensations. Condition (E) provides both aspiration/discharge and contact sensations.}}
	\end{table*}
	These results indicate that simultaneously providing both aspiration/discharge and contact sensations to users of an immersive micromanipulation system reduces the workload associated with micromanipulation.
	
	\begin{table}[!t]
		\caption{Experimental result of Q1: Ease of pipette movement.}
		\label{table:result_q1}
		\centering
		\begin{tabular}{c|cccc}
			\hline
			Condition & (B) & (C) & (D) & (E) \\
			Aspiration/discharge sensation & --- & \checkmark & --- & \checkmark \\
			Contact sensation & --- & --- & \checkmark & \checkmark \\ \hline
			Participant (a) & 6 & 6 & 6 & 6 \\ 
			Participant (b) & 6 & 6 & 6 & 6\\ 
			Participant (c) & 5 & 5 & 5 & 5 \\ 
			Participant (d) & 4 & 5 & 6 & 5 \\ 
			Participant (e) & 6 & 6 & 5 & 6\\
			Participant (f) & 5 & 6 & 6 & 5\\ \hline
			Median & 5.5 & 6.0 & 6.0 & 5.5\\ \hline
		\end{tabular}
	\end{table}
	\begin{table}[!t]
		\caption{Experimental result of Q2: Ease of aspiration and discharge.}
		\label{table:result_q2}
		\centering
		\begin{tabular}{c|cccc}
			\hline
			Condition & (B) & (C) & (D) & (E) \\
			Aspiration/discharge sensation & --- & \checkmark & --- & \checkmark \\
			Contact sensation & --- & --- & \checkmark & \checkmark \\ \hline
			Participant (a) & 3 & 5 & 6 & 6 \\ 
			Participant (b) & 2 & 5 & 5 & 6\\ 
			Participant (c) & 2 & 7 & 2 & 7 \\ 
			Participant (d) & 2 & 6 & 4 & 5 \\ 
			Participant (e) & 1 & 6 & 2 & 6\\
			Participant (f) & 2 & 6 & 3 & 6\\ \hline
			Median & 2.0 & 6.0$^{*}$ & 3.5 & 6.0$^{**}$\\ \hline
		\end{tabular}
		
		\vspace{2mm}
		\footnotesize{\hl{Significant difference compared with Condition (B) by Dunn's pairwise comparison with Bonferroni correction ($^{*}$~$p<0.05$, $^{**}$~$p<0.01$).}}
	\end{table}	
	\begin{table}[!t]
		\caption{Experimental result of Q3: Ease of holding a bead.}
		\label{table:result_q3}
		\centering
		\begin{tabular}{c|cccc}
			\hline
			Condition & (B) & (C) & (D) & (E) \\
			Aspiration/discharge sensation & --- & \checkmark & --- & \checkmark \\
			Contact sensation & --- & --- & \checkmark & \checkmark \\ \hline
			Participant (a) & 3 & 3 & 6 & 6 \\ 
			Participant (b) & 6 & 6 & 7 & 7\\ 
			Participant (c) & 5 & 6 & 7 & 7 \\ 
			Participant (d) & 4 & 4 & 5 & 5 \\ 
			Participant (e) & 2 & 2 & 2 & 7\\
			Participant (f) & 6 & 6 & 7 & 7\\ \hline
			Median & 4.5 & 5.0 & 6.5 & 7.0$^{*}$\\ \hline
		\end{tabular}
		
		\vspace{2mm}
		\footnotesize{\hl{Significant difference compared with Condition (B) by Dunn's pairwise comparison with Bonferroni correction ($^{*}$~$p<0.05$).}}
	\end{table}
	\begin{table}[!t]
		\caption{Experimental result of Q4: Ease of the experimental task.}
		\label{table:result_q4}
		\centering
		\begin{tabular}{c|cccc}
			\hline
			Condition & (B) & (C) & (D) & (E) \\
			Aspiration/discharge sensation & --- & \checkmark & --- & \checkmark \\
			Contact sensation & --- & --- & \checkmark & \checkmark \\ \hline
			Participant (a) & 5 & 6 & 6 & 6 \\ 
			Participant (b) & 5 & 5 & 5 & 7\\ 
			Participant (c) & 2 & 3 & 3 & 4 \\ 
			Participant (d) & 2 & 5 & 3 & 6 \\ 
			Participant (e) & 2 & 7 & 3 & 5\\
			Participant (f) & 2 & 6 & 6 & 6\\ \hline
			Median & 2.0 & 5.5 & 4.0 & 6.0$^{*}$\\ \hline
		\end{tabular}
		
		\vspace{2mm}
		\footnotesize{\hl{Significant difference compared with Condition (B) by Dunn's pairwise comparison with Bonferroni correction ($^{*}$~$p<0.05$).}}
	\end{table}
	Tables~\ref{table:result_q1}--\ref{table:result_q4} summarize the questionnaires results.
	Table~\ref{table:result_q1} shows the results for Q1 (Ease of pipette movement). The median scores range from 5.5 to 6.0 across all conditions, with no significant differences observed among conditions (B) through (E) \hl{($p = 0.58$, $W = 0.11$)}.
	Table~\ref{table:result_q2} shows the results for Q2 (Ease of aspiration and discharge). \hl{The Friedman test revealed a significant difference among the conditions ($p = 0.003$, $W = 0.78$). Post-hoc Dunn's pairwise comparisons with Bonferroni correction showed significant differences between Condition (B) (without any haptic sensations) and Conditions (C) and (E), both of which provided aspiration/discharge sensations ($p = 0.031$ and $p = 0.007$, respectively; differences in mean ranks $= 2.08$ and $2.42$).} The median scores are 2.0 for Condition (B), 6.0 for Condition (C), 3.5 for Condition (D), and 6.0 for Condition (E).
	Table~\ref{table:result_q3} presents the results for Q3 (Ease of holding a bead). \hl{The Friedman test revealed a significant difference among the conditions ($p = 0.001$, $W = 0.88$). A post-hoc Dunn's pairwise comparison with Bonferroni correction showed a significant difference between Conditions (B) and (E) (the latter provided both aspiration/discharge and contact sensations) ($p = 0.031$, difference in mean ranks $= 2.08$).} The median score increases from 4.5 in Condition (B) to 7.0 in Condition (E).
	Table~\ref{table:result_q4} presents the results for Q4 (Ease of the experimental task). \hl{The Friedman test revealed a significant difference among the conditions ($p = 0.004$, $W = 0.76$). A post-hoc Dunn's pairwise comparison with Bonferroni correction showed a significant difference between Conditions (B) and (E) ($p = 0.010$, difference in mean ranks $= 2.33$).} The median score improves from 2.0 in Condition (B) to 6.0 in Condition (E).
	These results indicate that providing aspiration and discharge sensations significantly improves users’ perceived ease in performing aspiration and discharge operations in the immersive micromanipulation system. 
	Furthermore, combining aspiration/discharge and contact sensations significantly enhances the user's perceived ease in holding a bead and completing the overall task.
	\fin{These differences remained significant under the conservative Bonferroni correction, which indicates that the significant results are robust despite the small sample.}
	
	\section{Discussion}
	\label{sec:discussion}
	The proposed immersive micromanipulation system reduced the task completion time for micromanipulation tasks.
	\hl{The immersive conditions (Conditions (B)--(E)) completed the task significantly faster than the conventional micromanipulation system (Condition (A)), which is consistent with our previous finding that immersive operation improves the speed of micromanipulation}~\cite{yokoe2022immersive}\hl{.}
	This improvement is likely due to the elimination of device-switching, achieved by integrating aspiration/discharge and pipette movement operations into a single interface within the immersive environment.
	Several previous studies have shown that reducing the number of operation interfaces contributes to improved operation speed in teleoperation~\cite{fang2025Effects,yokoe2025Automatica}. Among these approaches, hand gesture-based interfaces are highly intuitive and have been shown to significantly improve operational speed~\cite{zick2024Teleoperation,yeh2025Intuitive}.
	This result emphasizes the importance of interface reduction and hand gesture utilization in interface design.
	\hl{Therefore, the reduction in operation time achieved by the proposed system }\fin{may contribute}\hl{ to the quality of the operation by decreasing the exposure time of the target cell to stress. A direct evaluation of operation quality, such as cell damage and the number of operational errors using biological cells, can be pursued as future work.}
	
	\fin{Condition (B) serves as an internal visual-only baseline rather than a quantitative reproduction of the previous VR-based systems}~\cite{ammi2004Virtualized,mehrtash2012Humanassisted,bolopion2013Review,yokoe2022immersive}\hl{; thus, comparing it with the proposed haptic conditions (C)--(E) under identical settings provides a controlled }\fin{evaluation of the haptic sensations}\hl{. Adding both aspiration/discharge and contact sensations (Condition (E)) reduced the NASA-TLX WWL from 59.5 to 45.7 and improved usability (Q2--Q4), whereas the task completion time was maintained (38.9 s to 36.8 s). These results show that }\fin{the haptic sensations reduce}\hl{ the workload and }\fin{improve}\hl{ the usability while maintaining the operation speed.}
	
	Based on the NASA-TLX WWL scores, the haptic sensations of aspiration and discharge significantly contributed to workload reduction, enabling operators to accurately identify which operation (aspiration or discharge) was performed. 
	Likewise, presenting a haptic sensation resembling gripping the palm when a holding pipette contacts an oocyte may have contributed to the observed reduction in workload.
	Multiple Resource Theory suggests that different modalities do not share the same cognitive resource pool~\cite{wickens2002Multiple,wickens2008Multiple}. 
	Therefore, haptic feedback is expected to reduce visual cognitive load, which has been demonstrated through cognitive neuroscience experiments~\cite{marucci2021impact,du2024Sensory}.
	Multimodal interfaces incorporating haptics are important for reducing human cognitive load, and future interface design should consider how to effectively integrate multiple modalities. This study focuses only on vision and haptics; however, incorporating auditory feedback may further reduce cognitive load.
	The importance of multimodal interfaces incorporating haptics is supported by the other experimental results, which showed that Condition (E) (with both aspiration/discharge and contact sensations) significantly improved the scores for Q2 (Ease of aspiration/discharge), Q3 (Ease of holding), and Q4 (Ease of task performance) compared with Condition (B) (without any haptic sensations).
	
	\hl{The NASA-TLX WWL alone cannot clearly separate cognitive workload from physical workload. However, several observations indicate that the reduction was mainly cognitive and perceptual. Conditions (B)--(E) used the identical hand-tracking motion and differed only in the presented haptic information, which made the physical demand nearly the same across these conditions. Moreover, the improvements indicated by the questionnaire described above concern how easily the operator could perceive and perform the operations. These are perceptual and cognitive aspects, rather than physical effort. As described above through the Multiple Resource Theory, the haptic feedback offloaded the visual and cognitive channel, which explains the observed reduction in the WWL as a reduction in cognitive, rather than physical workload. This interpretation is supported by the NASA-TLX subscale ratings in Table~}\ref{table:result_subscale}\hl{, in which the Mental Demand subscale decreased the most from Condition (B) to Condition (E), whereas the physical demand subscale changed little.}
	
	The absence of significant differences in operation time among immersive environment conditions (B)--(E) may be because of the brief duration required to recognize aspiration/discharge and oocyte contact. Thus, although haptic sensations enhanced ease of recognition, \fin{no significant difference in task completion time was detected under the present statistical power, and a false negative cannot be excluded for this and the other non-significant results.}
	
	We used McKibben-based haptic sensation, which enabled users to intuitively recognize aspiration, discharge, and contact information, thereby successfully reducing cognitive load. However, other haptic sensations, such as vibrotactile feedback, could serve as alternatives. 
	Although other haptic sensations may reduce cognitive load, a critical consideration in selecting haptic modalities and algorithms is whether the sensations can intuitively convey aspiration, discharge, and contact information.
	
	Overall, these results demonstrate that immersive operation interface can shorten the operation time required for micromanipulations. Moreover, even when users perform unfamiliar operations in an immersive environment, presenting haptic sensations that are not normally perceivable does not increase workload or reduce usability, despite providing additional information. 
	Instead, haptic feedback contributes to a reduced workload and improved usability.
	
	\hl{Further studies in the future will evaluate the system with more participants, including experienced embryologists, and with biological cells. Experienced embryologists already have a strong internal model of the procedure; therefore, they are expected to benefit from the reduced interface switching and from the explicit presentation of the otherwise imperceptible aspiration, discharge, and contact states. This study used rigid microbeads, whereas biological cells such as oocytes have a deformable membrane surrounded by the zona pellucida. The contact-detection threshold and the gain $G$ in Equation~(}\ref{eq:target_angle}\hl{), which sets the injector rotation per grab angle, will, therefore, need to be recalibrated for biological cells because excessive injector rotation could apply excess force and damage the cell.}
	
	%The proposed approach of presenting imperceptible information through haptic feedback has significant implications for human--machine interaction. 
	%By augmenting visual feedback with artificial tactile sensations corresponding to operations that are visually difficult to perceive, operators can perform complex micromanipulation tasks more efficiently and with lower cognitive burden. 
	%This finding suggests that the design principle of ``augmenting imperceptible information" may apply to other domains where operators must infer system states from limited sensory information. 	
	
	\section{Conclusion}
	\label{sec:conclusion}
	We developed an immersive micromanipulation system that integrated pipette movement, aspiration, and discharge operations into a unified hand-tracking interface while providing McKibben-based haptic sensations. The proposed system generated haptic feedback representing aspiration, discharge, and contact states, eliminating the need for interface switching between pipette control and injector operations.
	
	Experimental evaluation with novice operators performing microbead manipulation tasks demonstrated that the immersive environment significantly reduced task completion time compared with conventional micromanipulation methods. Furthermore, the integration of haptic feedback significantly reduced cognitive workload and improved system usability, particularly for aspiration and discharge operations that produce only subtle visual cues.
	
	The following are the key conclusions from this work. 
	First, unifying pipette movement, aspiration, and discharge operations into a single hand-tracking interface eliminates interface switching and significantly reduces the task completion time in micromanipulation. 
	Second, McKibben-based fabric actuators successfully provide haptic sensations for imperceptible states, including aspiration, discharge, and contact detection. 
	Third, the proposed system significantly reduces cognitive workload and improves system usability for novice operators, particularly for operations that produce only subtle visual cues.
	
	Future work will focus on optimizing the artificial muscle alignment to improve haptic feedback quality and extending the system to perform other micromanipulation tasks by leveraging the capability of fabric actuators to provide multiple haptic sensations. 
	In addition, long-term evaluation studies with embryologists in clinical settings will provide insights into the practical applicability of the proposed system for ICSI training and procedure support.	
	
	\bibliography{saito_glove}
	\bibliographystyle{IEEEtran}
	
	\begin{IEEEbiography}[{\includegraphics[width=1in,height=1.25in,clip,keepaspectratio]{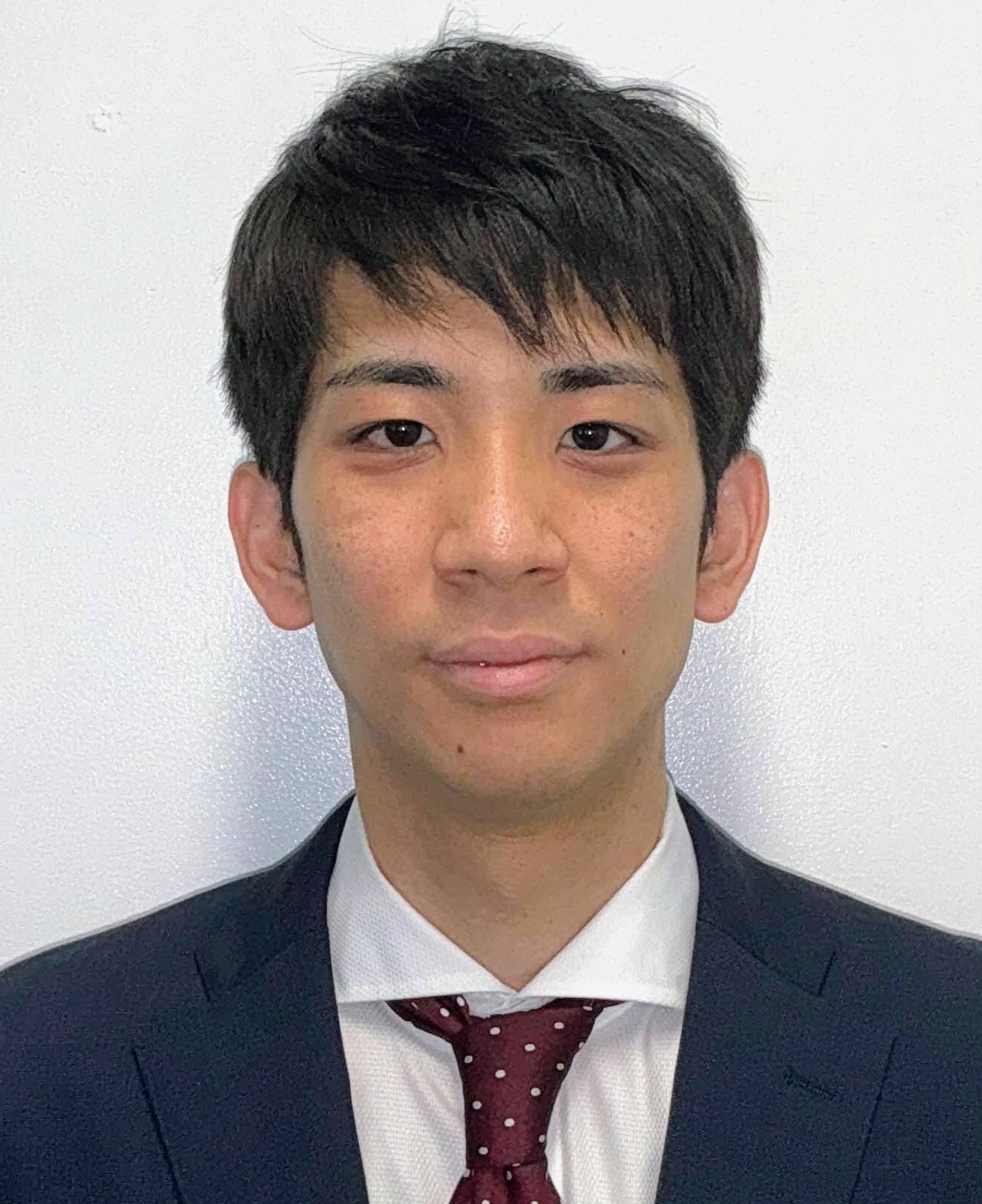}}]{Kenta Yokoe}
		(Member, IEEE) received the B.E. degree in mechanical engineering, the M.E., and the Ph.D. degrees in micro--nano mechanical science and engineering from Nagoya University, Nagoya, Japan, in 2021, 2023, and 2025, respectively. 
		He is currently an Assistant Professor at Nagoya University. His research interests include immersive systems, human interfaces, human--machine interaction, and haptic technology.
	\end{IEEEbiography}
	
	\begin{IEEEbiography}[{\includegraphics[width=1in,height=1.25in,clip,keepaspectratio]{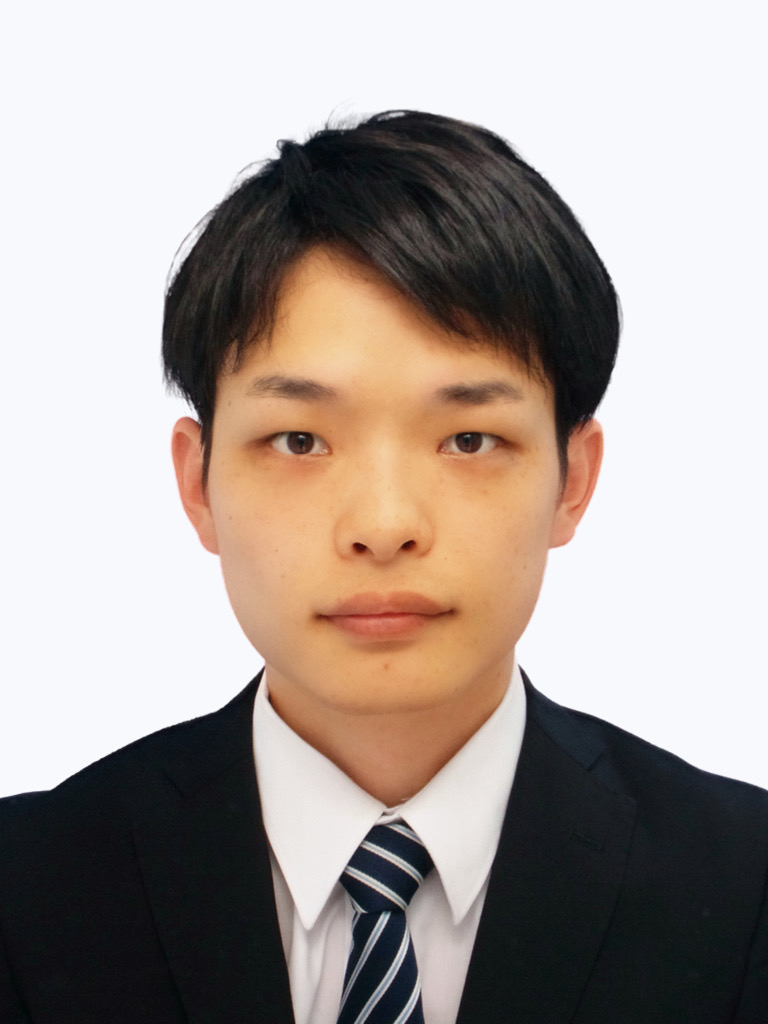}}]{Sumiwa Saito}
		received the B.E. degree in mechanical engineering and the M.E. degree in micro--nano mechanical science and engineering from Nagoya University, Nagoya, Japan, in 2022 and 2024, respectively. He is currently with Brother Industries, Ltd.
	\end{IEEEbiography}
	
	\begin{IEEEbiography}[{\includegraphics[width=1in,height=1.25in,clip,keepaspectratio]{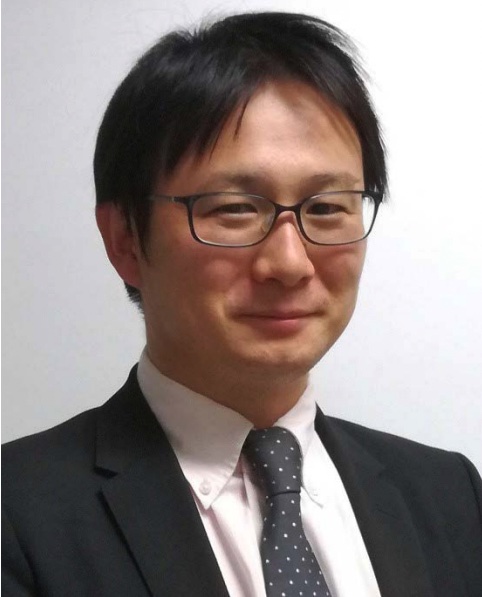}}]{Yuki Funabora}
		(Member, IEEE) received the B.E., M.E., and Ph.D. degrees in electrical engineering and computer science from Nagoya University, Nagoya, Japan, in 2007, 2009, and 2012, respectively. 
		In 2012, he was a Postdoctoral Researcher at the RIKEN Advanced Science Institute. 
		He was an Assistant Professor at Nagoya University from 2013 to 2021. 
		He was also a PRESTO Researcher at JST from 2018 to 2022. 
		He is currently an Associate Professor in the Department of Information and Communication Engineering at Nagoya University and a FOREST researcher at JST since 2021 and 2022, respectively. His research interests include human-cooperative robots, soft robotics, intelligent control, soft computing, and system design.
	\end{IEEEbiography}
	
	\begin{IEEEbiography}[{\includegraphics[width=1in,height=1.25in,clip,keepaspectratio]{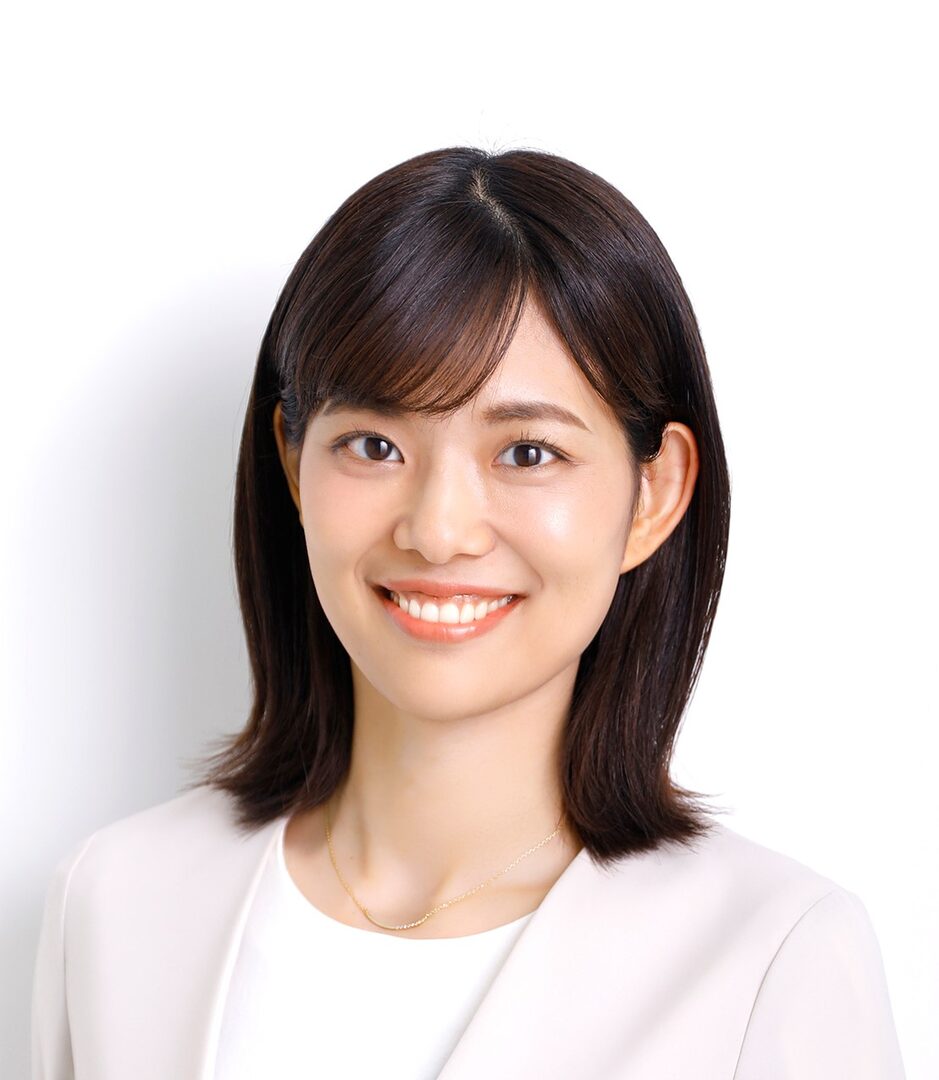}}]{Tomoko Isomura}
		received the B.Sc. from Waseda University in 2010, and both the M.Sc. and Ph.D. in Science from Kyoto University in 2012 and 2015, respectively. She has been an Associate Professor at the Graduate School of Informatics, Department of Cognitive and Psychological Sciences, Nagoya University, since 2020. Her research interests include cognitive, physiological, and developmental psychology.
	\end{IEEEbiography}
	
	\begin{IEEEbiography}[{\includegraphics[width=1in,height=1.25in,clip,keepaspectratio]{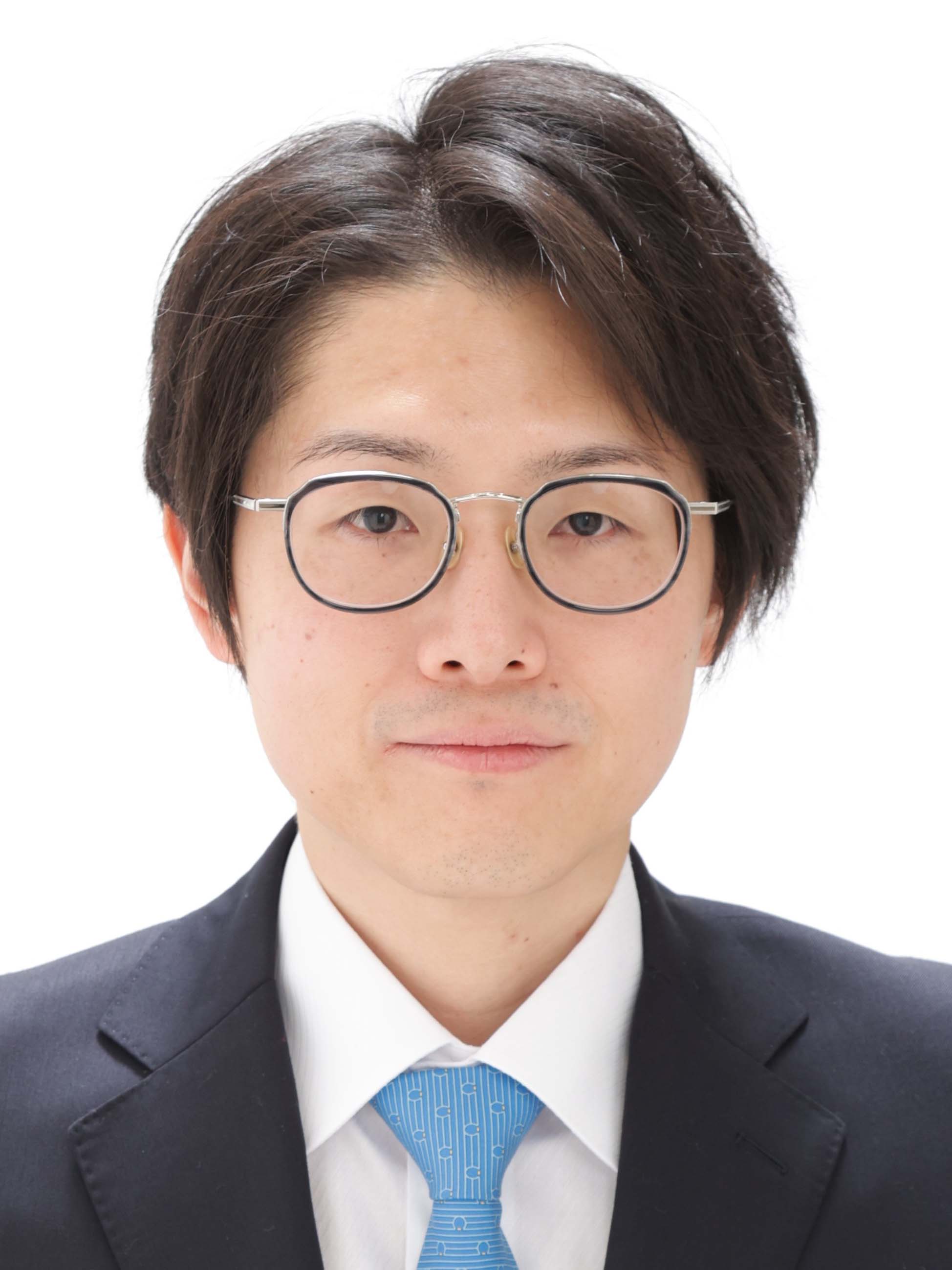}}]{Tadayoshi Aoyama}
		(Member, IEEE) received the B.E. degree in Mechanical Engineering, the M.E. degree in Mechanical Science and Engineering, and the Ph.D. degree in Micro--Nano Systems Engineering from Nagoya University, Nagoya, Japan, in 2007, 2009, and 2012, respectively.
		From 2012 to 2017, he was an Assistant Professor with Hiroshima University, and from 2017 to 2019, with Nagoya University. He then served as an Associate Professor with Nagoya University from 2019 to 2024. From 2018 to 2022, he was also a Researcher with the Precursory Research for Embryonic Science and Technology (PRESTO) program of the Japan Science and Technology Agency (JST).
		He is currently a Professor with the Department of Mechanical Systems Engineering, Nagoya University. His research interests include macro--micro interaction systems, human--AI cooperative interfaces, virtual and augmented reality, AI-based assistive technologies, micromanipulation, and medical robotics. 
	\end{IEEEbiography}
	
	\EOD
\end{document}